\documentclass[%
 aip,
 amsmath,amssymb,
 reprint,%
]{revtex4-1}

\usepackage{graphicx}
\usepackage{dcolumn}
\usepackage{bm}

\usepackage[utf8]{inputenc}
\usepackage[T1]{fontenc}
\usepackage{mathptmx}
\usepackage{etoolbox}
\usepackage{gensymb}
\usepackage{miller}
\usepackage{mathtools}
\usepackage{placeins}
\usepackage{dsfont}
\usepackage{braket}
\usepackage{xcolor}
\usepackage{siunitx}
\usepackage{ulem}

\makeatletter
\def\@email#1#2{%
 \endgroup
 \patchcmd{\titleblock@produce}
  {\frontmatter@RRAPformat}
  {\frontmatter@RRAPformat{\produce@RRAP{*#1\href{mailto:#2}{#2}}}\frontmatter@RRAPformat}
  {}{}
}%
\makeatother

\begin{document}


\title{Electronic band structure of superconducting KTaO$_3$ (111) interfaces} 



\author{Srijani Mallik}
\altaffiliation{these authors contributed equally to this work}
\affiliation{Unité Mixte de Physique, CNRS, Thales, Université Paris-Saclay, 91767 Palaiseau, France}
\email{srijani.mallik@cnrs-thales.fr}
\author{Börge Göbel}
\altaffiliation{these authors contributed equally to this work}
\affiliation{Institute of Physics, Martin-Luther-Universität Halle-Wittenberg, 06099 Halle, Germany}
\email{boerge.goebel@physik.uni-halle.de}
\author{Hugo Witt}
\altaffiliation{these authors contributed equally to this work}
\affiliation{Unité Mixte de Physique, CNRS, Thales, Université Paris-Saclay, 91767 Palaiseau, France}
\affiliation{Laboratoire de Physique et d’Etude des Matériaux, ESPCI Paris, Université PSL, CNRS, 75005, Paris, France}
\author{Luis M. Vicente-Arche}
\affiliation{Unité Mixte de Physique, CNRS, Thales, Université Paris-Saclay, 91767 Palaiseau, France}
\author{Sara Varotto}
\affiliation{Unité Mixte de Physique, CNRS, Thales, Université Paris-Saclay, 91767 Palaiseau, France}
\author{Julien Bréhin}
\affiliation{Unité Mixte de Physique, CNRS, Thales, Université Paris-Saclay, 91767 Palaiseau, France}
\author{Gerbold Ménard}
\affiliation{Laboratoire de Physique et d’Etude des Matériaux, ESPCI Paris, Université PSL, CNRS, 75005, Paris, France}
\author{Guilhem Sa\"iz}
\affiliation{Laboratoire de Physique et d’Etude des Matériaux, ESPCI Paris, Université PSL, CNRS, 75005, Paris, France}
\author{Dyhia Tamsaout}
\affiliation{Unité Mixte de Physique, CNRS, Thales, Université Paris-Saclay, 91767 Palaiseau, France}
\author{Andrés Felipe Santander-Syro}
\affiliation{Institut des Sciences Moléculaires d'Orsay, CNRS, Université Paris-Saclay, 91405 Orsay, France}
\author{Franck Fortuna}
\affiliation{Institut des Sciences Moléculaires d'Orsay, CNRS, Université Paris-Saclay, 91405 Orsay, France}
\author{François Bertran}
\affiliation{SOLEIL synchrotron, L’Orme des Merisiers, Départementale 128, F-91190 Saint-Aubin, France}
\author{Patrick Le Fèvre}
\affiliation{SOLEIL synchrotron, L’Orme des Merisiers, Départementale 128, F-91190 Saint-Aubin, France}
\author{Julien Rault}
\affiliation{SOLEIL synchrotron, L’Orme des Merisiers, Départementale 128, F-91190 Saint-Aubin, France}
\author{Isabella Boventer}
\affiliation{Unité Mixte de Physique, CNRS, Thales, Université Paris-Saclay, 91767 Palaiseau, France}
\author{Ingrid Mertig}
\affiliation{Institute of Physics, Martin-Luther-Universität Halle-Wittenberg, 06099 Halle, Germany}
\author{Agnès Barthélémy}
\affiliation{Unité Mixte de Physique, CNRS, Thales, Université Paris-Saclay, 91767 Palaiseau, France}
\author{Nicolas Bergeal}
\affiliation{Laboratoire de Physique et d’Etude des Matériaux, ESPCI Paris, Université PSL, CNRS, 75005, Paris, France}
\author{Annika Johansson}
\affiliation{Max Planck Institute of Microstructure Physics, Weinberg 2, 06120 Halle, Germany}
\email{annika.johansson@mpi-halle.mpg.de}
\author{Manuel Bibes}
\affiliation{Unité Mixte de Physique, CNRS, Thales, Université Paris-Saclay, 91767 Palaiseau, France}
\email{manuel.bibes@cnrs-thales.fr}


\date{\today}

\begin{abstract}
Two-dimensional electron gases (2DEGs) based on KTaO$_3$ are emerging as a promising platform for spin-orbitronics due to their high Rashba spin-orbit coupling (SOC) and gate-voltage tunability. The recent discovery of a superconducting state in KTaO$_3$ 2DEGs now expands their potential towards topological superconductivity. Although the band structure of KTaO$_3$ surfaces of various crystallographic orientations has already been mapped using angle-resolved photoemission spectroscopy (ARPES), this is not the case for superconducting KTaO$_3$ 2DEGs. Here, we reveal the electronic structure of superconducting 2DEGs based on KTaO$_3$ (111) single crystals through ARPES measurements.  We fit the data with a tight-binding model and compute the associated spin textures to bring insight into the SOC-driven physics of this fascinating system.
\end{abstract}


\pacs{}

\maketitle 

\section{Introduction}

Oxide interfaces can harbour exotic phases of condensed matter, often absent in the interface constituents of their bulk form\cite{hwang_emergent_2012,trier_oxide_2022}. A paradigmatic example is the SrTiO$_3$ (STO) two-dimensional electron gas (2DEG) that forms when STO is interfaced with epitaxial oxides such as LaAlO$_3$ (LAO)\cite{ohtomo_high-mobility_2004} or through local redox processes occurring at the interface when STO is covered with a reactive metal such as Al or Ta\cite{rodel_universal_2016,vicente-arche_metal_2021,grezes_non-volatile_2023}. While some properties of STO 2DEGs can be found in bulk STO, including high mobility transport\cite{frederikse_electronic_1964} and low temperature superconductivity\cite{schooley_superconductivity_1964}, these are often superior or of a different nature in STO 2DEGs. For instance, mobilities exceeding 10$^5$ cm$^2$V$^{-1}$s$^{-1}$ have been reported in STO-based 2DEGs\cite{chen_high-mobility_2013}, while the record value for bulk STO is $\sim$10$^4$ cm$^2$V$^{-1}$s$^{-1}$ in bulk STO, and superconductivity in 2DEGs is two-dimensional \cite{reyren_superconducting_2007} and highly tunable by electrostatic gating\cite{caviglia_electric_2008}. Moreover, 2DEGs also display unique features, such as Rashba spin-orbit coupling (SOC) \cite{caviglia_tunable_2010}, that only arises in environments with broken inversion symmetry (e.g. surfaces or interfaces), and intrinsic signatures of two-dimensional transport such as the quantum Hall effect\cite{trier_quantization_2016}. 

Key insight into the physics of STO 2DEGs was brought through angle-resolved photoemission spectroscopy (ARPES) by mapping their electronic structure. Such measurements revealed band splittings and sub-bands arising from quantum confinement\cite{santander-syro_two-dimensional_2011,berner_direct_2013,mckeown_walker_absence_2016,vaz_determining_2020}, which are absent in the bulk\cite{mattheiss_energy_1972}. Most ARPES results have been collected on (001)-oriented STO 2DEGs and have shown the coexistence of bands having a d$_{xy}$ orbital character with low effective mass and bands having a d$_{xz/yz}$ character with a higher effective mass. Orbital mixing due to the reduced symmetry produces avoided crossings where SOC-related effects are enhanced\cite{vaz_determining_2020}. Such studies dramatically advanced the understanding of the superconducting phase diagram (with heavier bands playing a key role) and of spin-charge interconversion phenomena driven by the Rashba SOC\cite{lesne_highly_2016,chauleau_efficient_2016,vaz_determining_2020,vaz_mapping_2019,ohya_efficient_2020}. 
The electronic structure of STO 2DEGs grown along other crystal orientations, namely (110) and (111), has also been mapped by ARPES\cite{rodel_orientational_2014,mckeown_walker_control_2014}. These experiments helped to understand the differences in their physical response, notably regarding the dependence of superconductivity on Fermi energy. For 2DEGs oriented along (110) and (111) directions, this dependence is much less pronounced due to their reduced orbital splitting (see e.g. Refs. \cite{singh_gap_2019,jouan_multiband_2022}).  

Similar to STO, KTaO$_3$ (KTO) is a quantum paraelectric that in the bulk becomes metallic upon minute electron doping \cite{frederikse_electronic_1964,wemple_transport_1965}. This prompted the exploration of KTO interfaces a few years after the discovery of STO 2DEGs\cite{king_subband_2012,santander-syro_orbital_2012,zou_latio_2015,gupta_ktao3_2022}. Aside from measurements on ionic-liquid gated crystals showing an ultralow critical temperature $T_c$ of 40 mK\cite{ueno_discovery_2011}, initially, no superconductivity was found in KTO 2DEGs, consistent with the absence of superconducting properties in bulk KTO\cite{thompson_very_1982}. At first, efforts on KTO 2DEGs were concentrated on the (001) orientation, and one notable finding was the presence of a large Rashba SOC through weak antilocalization or spin-charge interconversion experiments\cite{vicentearche_spincharge_2021}, where the Rashba coefficient was found to be 5-10 times larger than in STO, as expected from the heavier mass of Ta compared to Ti. The sizeable band splitting induced by Rashba SOC was sufficiently large to be directly imaged by ARPES\cite{varotto_direct_2022}, which also revealed bands with well-defined orbital character, akin to the situation in STO 2DEGs. 

In this context, the discovery of superconductivity with a $T_c$ up to 2.2 K in (111)-oriented KTO 2DEGs came as a big surprise\cite{liu_two-dimensional_2021}. This finding was soon confirmed by different groups, see e.g. Refs. \cite{mallik_superfluid_2022,arnault_anisotropic_2023,qiao_gate_2021}. Superconductivity was also found up to about 1 K in (110)-oriented KTO 2DEGs\cite{hua_tunable_2022}. To this day, the reasons for this huge orientational dependence -- at odds with the situation in STO 2DEGs -- have not been clarified, even if some  theoretical proposals have recently emerged, discussing the role of electron-phonon coupling or of the Rashba SOC in driving the superconductivity \cite{gastiasoro_theory_2022,liu_tunable_2023,esswein_first-principles_2023}. To resolve this conundrum, high quality ARPES measurements to precisely map and understand the electronic structure of superconducting KTO 2DEGS are needed. A few years before the discovery of superconductivity in KTO (111), two studies reported the band structure of KTO (111) surfaces using ARPES \cite{bareille_two-dimensional_2015,bruno_band_2019}. The band structure of KTO (110) 2DEGs has also been reported not long ago\cite{Mart_nez_2023}. Yet, whether these samples were superconducting or not remains unknown. Recently Chen \textit{et al}. have reported ARPES data of KTO (001), superconducting KTO (110) and (111) measured using soft x-rays with 1000 eV photon energy. However, due to this high photon energy, it was not possible to resolve the exact band structure near the Fermi energy\cite{chen2023orientationdependent}. 

In this paper, we report the electronic structure of KTO~(111) 2DEGs by ARPES measured with high energy resolution using a photon energy of 31 eV. The samples are prepared by depositing a few \AA~of Eu, triggering a redox reaction to form the 2DEG. They are superconducting with an anisotropic $T_c$ of 0.6 and 0.8 K, depending on the direction along which the current is applied. The dispersion curves and Fermi surfaces obtained by ARPES were fitted with four spin-resolved band pairs. We calculate the band-resolved spin textures as a function of energy and discuss the results in light of future spin-charge interconversion experiments and the origin of superconductivity in this system.

\section{Methods}
\subsection{Sample preparation}
Single crystalline KTO~(111) substrates from MTI Corporation were pre-annealed at 300 $\degree$C for two hour in ultra-high vacuum inside a molecular beam epitaxy (MBE) chamber. We grew 3 \AA~of Eu at 300 $\degree$C using a Knudsen cell heated to 500 $\degree$C at a growth rate of 0.043 \AA .s$^{-1}$ and at a deposition pressure of 1.2 $\times$ 10$^{-9}$ mbar. Subsequently, the sample was transferred in-situ to the ARPES chamber.

For ex-situ transport measurement, another sample with 3 \AA~Eu was prepared in the exact same deposition conditions. In order to protect the 2DEG at the Eu/KTO~(111) interface from oxidation from the air, we capped the sample in-situ with 2.1 nm of Al grown at room temperature using a Knudsen cell heated to 1000 $\degree$C at a growth rate of 0.092 \AA .s$^{-1}$. Thus, the Al overlayer was spontaneously and completely oxidized into AlO$_x$ when exposed to the air prior to the transport measurements. Both samples were grown on two pieces cleaved from the same KTO~(111) substrate.


\subsection{Angular resolved photoemission spectroscopy}

High-resolution angular resolved photoemission spectroscopy (ARPES) measurements were performed at the Cassiopée beamline of Synchrotron SOLEIL (France). This beamline is equipped with a Scienta R4000 hemispherical electron energy analyser. All the measurements were performed at 15 K (sample space cooled by liquid He) in order to minimize the thermal broadening of the spectral lines. The energy and angular resolution were 15 meV and $<$ 0.25$\degree$, respectively. Data presented in the manuscript were collected with linear horizontally (LH) polarized photons with a photon energy of 31 eV. The collected data were normalized by taking the second derivative of the intensity background of the electron analyzer and smoothed using an averaging filter.

\subsection{Tight-binding model}
Our tight-binding (TB) model considers 36 bands (18 band pairs). One unit cell considers three of each of the 5\textit{d} Ta $t_{2g}$ orbitals ($d_{xy}$, $d_{xz}$, $d_{yz}$) giving rise to the basis 
\begin{align}\label{eq:basis}
\{ \ket{d_{xy1\uparrow}},\ket{d_{xy1\downarrow}} ,\ket{d_{yz1\uparrow}},\ket{d_{yz1\downarrow}} ,\ket{d_{zx1\uparrow}},\ket{d_{zx1\downarrow}}&,\notag\\\ket{d_{xy2\uparrow}},\ket{d_{xy2\downarrow}} ,\ket{d_{yz2\uparrow}},\ket{d_{yz2\downarrow}} ,\ket{d_{zx2\uparrow}},\ket{d_{zx2\downarrow}}&,\notag\\\ket{d_{xy3\uparrow}},\ket{d_{xy3\downarrow}} ,\ket{d_{yz3\uparrow}},\ket{d_{yz3\downarrow}} ,\ket{d_{zx3\uparrow}},\ket{d_{zx3\downarrow}} &\}.
\end{align}
We add another copy with different parameters to reproduce the subbands arising due to the quantum confinement of electrons near the surface.
We diagonalize the tight-binding Hamiltonian
\begin{align}\label{eq:Hamiltonian}
H=H_\mathrm{hop}+H_\mathrm{SOC}+H_\mathrm{mix}
\end{align}
to describe the electrons gas that is confined at the interface of KTO~(111) with a lattice constant $a=4.0~\,\mathrm{\mathring{A}}$ formed by the Ta atoms.

For the hopping term we consider 
\begin{widetext}
\begin{align}
H_\mathrm{hop}=&\begin{psmallmatrix}
\Delta \epsilon &0&0&h_{\pi}^{-x}+h_{\pi}^{-y}+h_{\delta}^{-z}&0&0&0&0&0\\
0&\Delta \epsilon &0&0&h_{\delta}^{-x}+h_{\pi}^{-y}+h_{\pi}^{-z}&0&0&0&0\\
0&0&\Delta \epsilon &0&0&h_{\pi}^{-x}+h_{\delta}^{-y}+h_{\pi}^{-z}&0&0&0\\
h_{\pi}^x+h_{\pi}^y+h_{\delta}^z&0&0&\Delta \epsilon &0&0&h_{\pi}^{-x}+h_{\pi}^{-y}+h_{\delta}^{-z}&0&0\\
0&h_{\delta}^x+h_{\pi}^y+h_{\pi}^z&0&0&\Delta \epsilon &0&0&h_{\delta}^{-x}+h_{\pi}^{-y}+h_{\pi}^{-z}&0\\
0&0&h_{\pi}^x+h_{\delta}^y+h_{\pi}^z&0&0&\Delta \epsilon &0&0&h_{\pi}^{-x}+h_{\delta}^{-y}+h_{\pi}^{-z}\\
0&0&0&h_{\pi}^x+h_{\pi}^y+h_{\delta}^z&0&0&\Delta \epsilon &0&0\\
0&0&0&0&h_{\delta}^x+h_{\pi}^y+h_{\pi}^z&0&0&\Delta \epsilon &0\\
0&0&0&0&0&h_{\pi}^x+h_{\delta}^y+h_{\pi}^z&0&0&\Delta \epsilon
\end{psmallmatrix}\notag\\
&\bigotimes
\begin{pmatrix} 1&0\\0&1 \end{pmatrix}
\end{align}
\end{widetext}
with $h_{\alpha}^{j}=t_{\alpha}e^{i a k_j}$ indicating the nearest-neighbor $\alpha=\{\pi,\delta\}$ hopping of the $\{d_{xy},d_{yz},d_{zx}\}$ orbitals along $\{x,y,z\}$ according to the Slater-Koster rules. The linearly independent hopping amplitudes are $t_{\pi}=-0.65~\,\mathrm{eV}$, $t_{\delta}=-0.05~\,\mathrm{eV}$. Note, that the $\sigma$ hopping is not relevant for nearest-neighbor hoppings considering only the $t_{2g}$ orbitals. The onsite energies are $\Delta \epsilon=1.946~\,\mathrm{eV}$ (and $\Delta \epsilon=2.011~\,\mathrm{eV}$ for the subbands) causing a shift of the bands in energy.

The gradient potential at the interface displaces the oxygens' $p$ orbitals away from the bond connecting two neighboring Ta atoms. Hopping terms that are forbidden in bulk are now allowed by the Slater-Koster rules. By analogy with KTO~(001) 2DEGs, discussed in Ref.~\cite{varotto_direct_2022}, hopping between two neighboring Ta $d$ orbitals, via an intermediate hopping to an oxygen $p$ orbital, causes an effective hopping network. The hopping is asymmetric and gives rise to orbital mixing terms
\begin{widetext}
\begin{align}
H_\mathrm{mix}=&g\begin{psmallmatrix}
0&0&0&0&e^{-iak_z}-e^{-iak_x}&e^{-iak_z}-e^{-iak_y}&0&0&0\\
0&0&0&e^{-iak_x}-e^{-iak_z}&0&e^{-iak_x}-e^{-iak_y}&0&0&0\\
0&0&0&e^{-iak_y}-e^{-iak_z}&e^{-iak_y}-e^{-iak_x}&0&0&0&0\\
0&e^{iak_x}-e^{iak_z}&e^{iak_y}-e^{iak_z}&0&0&0&0&e^{-iak_z}-e^{-iak_x}&e^{-iak_z}-e^{-iak_y}\\
e^{iak_z}-e^{iak_x}&0&e^{iak_y}-e^{iak_x}&0&0&0&e^{-iak_x}-e^{-iak_z}&0&e^{-iak_x}-e^{-iak_y}\\
e^{iak_z}-e^{iak_y}&e^{iak_x}-e^{iak_y}&0&0&0&0&e^{-iak_y}-e^{-iak_z}&e^{-iak_y}-e^{-iak_x}&0\\
0&0&0&0&e^{iak_x}-e^{iak_z}&e^{iak_y}-e^{iak_z}&0&0&0\\
0&0&0&e^{iak_z}-e^{iak_x}&0&e^{iak_y}-e^{iak_x}&0&0&0\\
0&0&0&e^{iak_z}-e^{iak_y}&e^{iak_x}-e^{iak_y}&0&0&0&0
\end{psmallmatrix}\notag\\
&\bigotimes
\begin{pmatrix} 1&0\\0&1 \end{pmatrix}
\end{align}
\end{widetext}
The amplitude is $g=3~\,\mathrm{meV}$ (and $g=1~\,\mathrm{meV}$ for the subbands).

$H_\mathrm{SOC}$ with $\lambda=0.4~\,\mathrm{eV}$ describes the on-site spin-orbit coupling which mixes the different spin orientations and orbitals. It is the same for each lattice site and reads
\begin{align}
H_\mathrm{SOC}=\frac{\lambda}{3}\begin{pmatrix}
0&0&0&1&0&-i\\
0&0&-1&0&-i&0\\
0&-1&0&0&i&0\\
1&0&0&0&0&-i\\
0&i&-i&0&0&0\\
i&0&0&i&0&0
\end{pmatrix}
\end{align}
in the basis $\{ \ket{d_{xy\uparrow}},\ket{d_{xy\downarrow}} ,\ket{d_{yz\uparrow}},\ket{d_{yz\downarrow}} ,\ket{d_{zx\uparrow}},\ket{d_{zx\downarrow}} \}$.

\subsection{Transport measurements}
The transport measurements were performed in DC with the current flowing along the \hkl[1-10] and \hkl[-1-12] directions of the sample inside a dilution refrigerator with a base temperature of 25 mK. The sample holder is thermally decoupled from the mixing chamber plate to allow a continuous temperature sweep from 25 mK to 2 K.  

\section{Results and Discussion}
\subsection{Structural and transport properties}
\begin{figure}
\centering
\includegraphics[width=0.8\columnwidth]{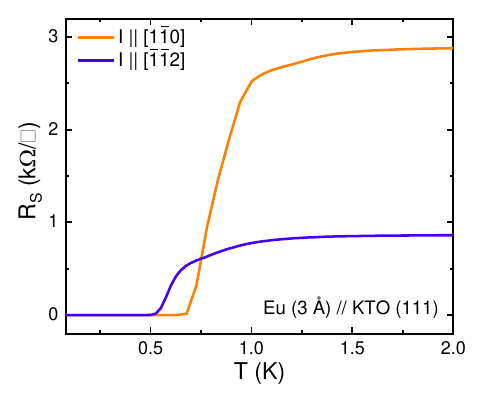}
\caption{\label{LEED-RT} \textbf{Transport measurement.} Temperature dependence of the sheet resistance measured with current parallel to \hkl[1-10] (orange) and \hkl[-1-12] (blue) directions where the superconducting transitions are observed at temperatures $T_C$ = 0.82 and 0.58 K, respectively.}
\end{figure}

Fig. \ref{LEED-RT} exhibits the temperature dependence of the sheet resistance of the sample capped with 2 nm of Al. It shows a metallic behavior from 300 K to $\sim$ 1 K (see Supplementary Material Fig.S1(a)) and then undergoes a superconducting transition. The orange and blue lines in Fig. \ref{LEED-RT} correspond to the data measured with the current along \hkl[1-10] and \hkl[-1-12] directions, respectively.  The difference in the resistance along \hkl[-1-12] and \hkl[1-10] directions at low temperatures suggests the presence of a strong in-plane anisotropy as demonstrated by previous studies \cite{liu_two-dimensional_2021}. The sample undergoes a superconducting transition along both directions with $T_c$ (middle of the transition) $\sim$ 0.82 K and 0.58~K. The magnetoresistance and Hall measurements at 2 K were carried out to estimate the carrier density and mobility (see Supplementary Material for detailed discussion). The mobilities were quite different along \hkl[-1-12] and \hkl[1-10], amounting to 111 cm$^2$V$^{-1}$s$^{-1}$ and 33 cm$^2$V$^{-1}$s$^{-1}$, respectively. This hints at an anisotropy in the band structure between the $\Gamma -\text{M}$ and $\Gamma -\text{K}$  directions.   

\subsection{Band structure of Eu/KTO~(111) 2DEG}

\begin{figure*}
\centering
\includegraphics[width=1.8\columnwidth]{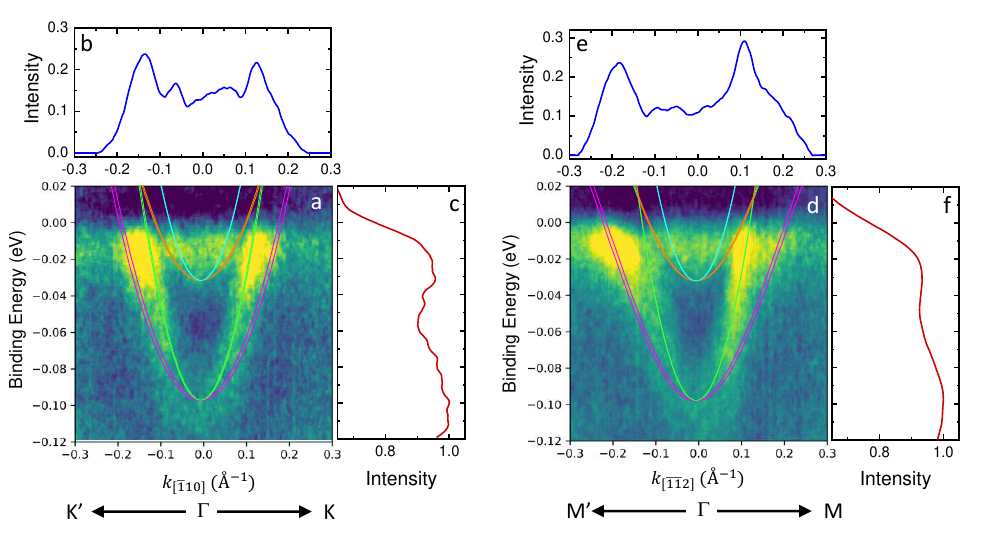}
\caption{\label{dispersion} \textbf{Electronic band structure of KTO~(111) 2DEG.} Band dispersion of Eu/KTO (111) 2DEG measured by ARPES along high symmetry directions (a) $\Gamma$ - K and (d) $\Gamma$ - M. The tight-binding fits of the band structure are overlaid to the data where a specific color is associated with each pair of bands. Momentum distribution curves (MDC) at $E_F$ are shown in (b) for $\Gamma$ - K and in (e) for $\Gamma$ - M directions. Energy distribution curves (EDC) at $\Gamma$ are shown in (c) and (f).}
\end{figure*} 

\begin{figure*}
\centering
\includegraphics[width=1.8\columnwidth]{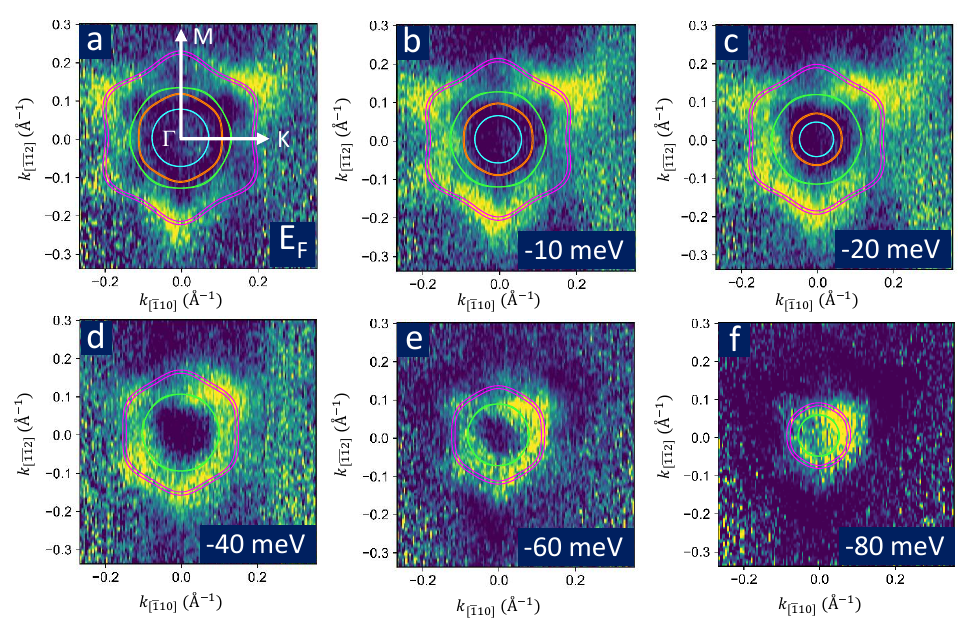}
\caption{\label{FS} \textbf{Fermi surfaces.} Constant energy maps with their corresponding tight-binding fit at (a) Fermi energy ($E_\mathrm{F}$), (b) 10 meV, (c) 20 meV, (d) 40 meV, (e) 60 mev, and (f) 80 meV below the Fermi level. The high symmetry points ($\Gamma$, M, K) and their directions are shown with white arrows on (a).}
\end{figure*}

The band dispersion of the Eu(3 \AA)/KTO~(111) sample was measured along the $\Gamma -\text{K}$ and $\Gamma -\text{M}$ directions with a photon energy of 31 eV at normal emission, see Fig. \ref{dispersion}(a) and (d). The data clearly indicate an anisotropic electronic structure. By performing a manual photon energy scan we also found photo emission intensity close to 106 eV of photon energy. This photon energy is similar to the one reported by Bruno \textit{et al}.\cite{bruno_band_2019}. However, we found the highest photo emission intensity at 31 eV and thus in the following we show the band structures measured using 31 eV of photon energy. The sample is superconducting but the ARPES measurements were carried out at the normal state at 15 K. Tight-binding fits of the band structure are overlaid on the ARPES dispersion data. Out of the eighteen pairs of bands in our TB model, four are visible in the measurement window. Similar to our previous report on KTO~(001) \cite{varotto_direct_2022}, we have associated four different colors for each pair of bands. The pink, green, and orange band pairs arise from the t$_{2g}$ orbitals whereas the cyan band pair is the contribution from another copy of the subband arising due to quantum confinement of the 2DEG. We have also considered the mixing of Ta 5d orbitals and the spin-orbit coupling. In Ref. \cite{varotto_direct_2022}, we could easily identify the orbital character of each band pair. However, unlike in KTO~(001) 2DEGs, all three $t_{2g}$ orbitals are highly hybridized in the case of KTO~(111) 2DEGs and therefore, it is not possible to ascribe a specific orbital character to each band pair, as also discussed in Ref. \cite{bruno_band_2019}. However, the contribution of orbitals with a given character depends on the in-plane direction. This can be inferred from the clear asymmetry observed in the contrast of the Fig. \ref{dispersion}(d) where the pink band pairs are better visible in the data along $\Gamma$ - M$'$ direction whereas the green band pair is better visible along $\Gamma - \text{M}$ direction. The energy distribution curves (EDC) at \textit{k} = 0 (Fig. 2(c and f)) and momentum distribution curves (MDC) at E$_F$ (Fig. 2(b and e)) corresponding to the dispersion curves for both $\Gamma$K and $\Gamma$M directions are analyzed to determine the quality of the TB fits with respect to the experimental observation of the band structure. The detailed analysis is discussed in the supplementary material. Further, the pink band pair in Fig. \ref{dispersion}(d) appears to be heavier in the raw data around \textit{k} = $\pm$ 0.2 \AA$^{-1}$ in comparison to the fit. An explanation for the apparent enhancement of the effective mass of the pink band pair could be an electron-phonon interaction that causes parabolic bands to look kinked below a certain binding energy. This effect has been observed previously in STO 2DEGs \cite{King2014NatureCommun}. The disagreement between the fit and the data occurs because our model does not consider electron-phonon coupling. From the TB calculation, we can estimate the effective mass. We find that the cyan and green band pairs exhibit a lower effective mass of 0.67 $m_e$ at the $\Gamma$ point in comparison to the orange and pink band pairs with a higher effective mass of 1.45 $m_e$ where $m_e$ is the electron mass. In general, the effective masses are $k$-dependent and anisotropic. The Rashba splittings near the $\Gamma$ point have also been calculated from the TB model. For the pink and orange band pairs, the splitting is isotropic near $\Gamma$ with values of $\alpha_R$ = 20 meV.\AA$^{-1}$ and 8 meV.\AA$^{-1}$, respectively. For the green and cyan bands the splitting is anisotropic. It is zero along $\Gamma -\text{M}$ and has values of $\alpha_R$ = 14 meV.\AA$^{-1}$ and 4 meV.\AA$^{-1}$ along $\Gamma -\text{K}$, respectively. The $\alpha_R$ values obtained for KTO~(111) 2DEGs are comparatively smaller with respect to the high $\alpha_R$ of $\sim$ 300 meV.\AA$^{-1}$ observed for the heavy band pair in KTO~(001) 2DEGs \cite{varotto_direct_2022}. A similar reduction in $\alpha_R$ was observed in the case of the SrTiO$_3$ (111)-based 2DEGs in comparison to the (001)-based ones \cite{PhysRevLett.119.237002}. We have not performed photon energy scans to prove the two-dimensional nature of the Eu/KTO (111) electron gas. However, from the dispersion measurements we observe sub-bands which are indicative of quantum confinement and consistent with a 2DEG.

\begin{figure*}
\centering
\includegraphics[width=2.0\columnwidth]{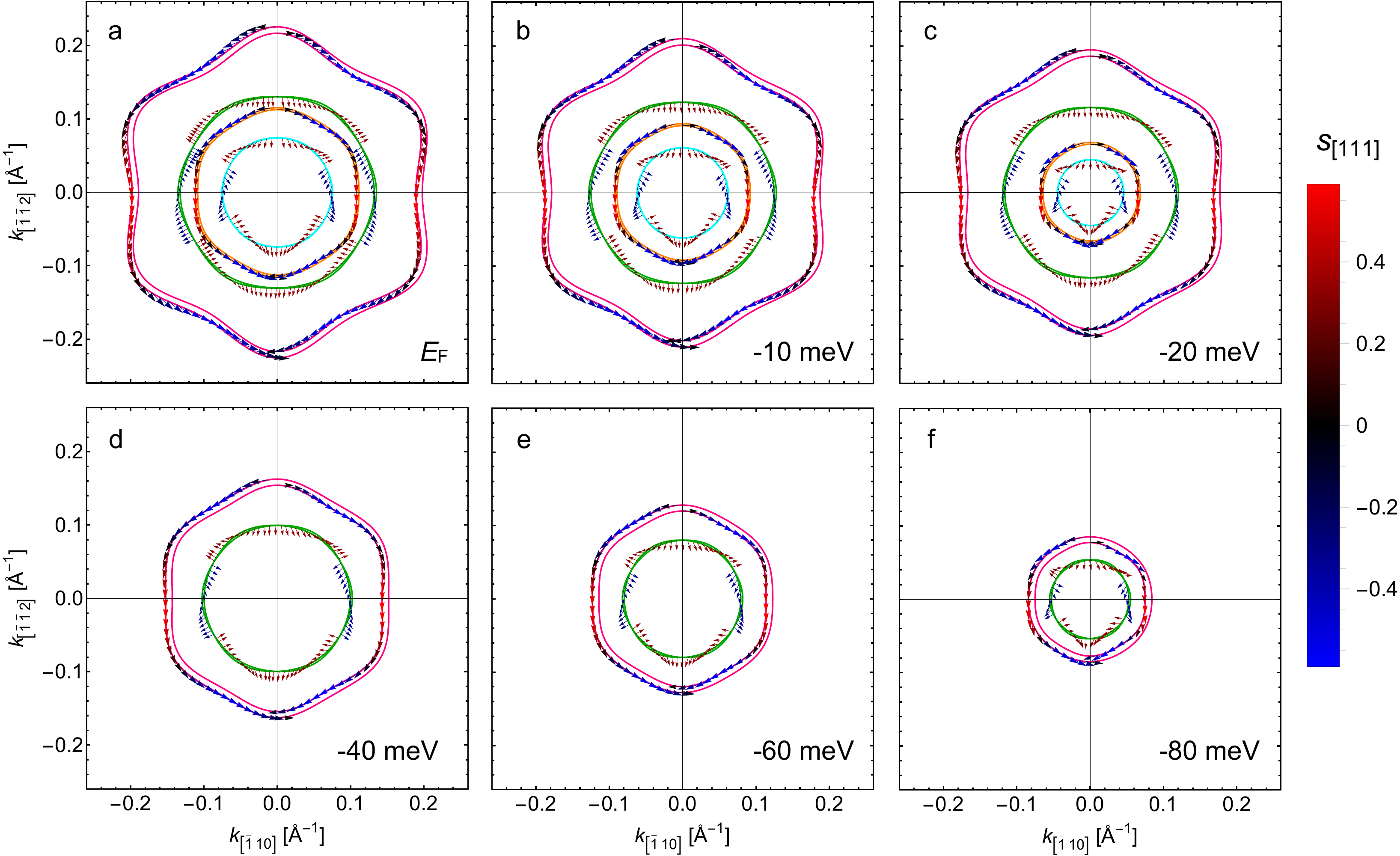}
\caption{\label{spin} \textbf{Spin textures.} The fitted iso-energy lines from Fig. 3 (same colors) with arrows indicating the spin texture. The arrow orientation and size correspond to the in-plane component. Their color indicates the out-of-plane component with red along \hkl[111], and blue along \hkl[-1-1-1]). In each panel, the left (right) half shows the spins of the outer (inner) band of each band pair. The spins of the green and cyan bands reverse at some points of the Fermi lines as the inner and outer bands intersect.}
\end{figure*}

Fig. \ref{FS} shows constant energy maps and their corresponding TB fits for different binding energies. We observe a star-shaped Fermi surface centered at $\Gamma$ with three concentric FSs. The star-shaped FS fitted by the pink band pair displays a major diameter of $\sim$ 0.47 \AA$^{-1}$ along the $\Gamma - \text{M}$ direction and a minor diameter of $\sim$ 0.35 \AA$^{-1}$ along the $\Gamma - \text{K}$ direction. Although the general star shape of the outermost FS matches with the one reported by Bruno \textit{et al.} \cite{bruno_band_2019}, the ratio between the major and minor diameters differs, suggesting a slightly different shape of the FS in the case of an MBE-grown KTO~(111) based 2DEG compared to a free KTO~(111) surface. At $E_F$ = 0, the TB fits suggest star-shaped FSs for the pink and orange band pair whereas the green and cyan band pairs display hexagonal and circular-shaped FSs, respectively. The total carrier density calculated from the TB model is 12.5$\times$10$^{13}$ cm$^{-2}$. This is in good agreement with the carrier density (11.4$\times10^{13}$ cm$^{-2}$) obtained from the first transport measurements performed just after the ARPES measurements (see Supplementary Material for detailed discussion). Bruno \textit{et al.} discussed in their TB model that each branch of the outermost star-shaped FS can possess different orbital characteristics \cite{bruno_band_2019}. Our measurements [Fig. \ref{FS}(a-c)] with LH polarization reveal the branches having $d_{xz}$ and $d_{xy}$ orbital characters and limit the possibility to visualize the branches possessing $d_{yz}$ orbital characteristics, which is consistent with Ref. \cite{bruno_band_2019}.

\subsection{Spin textures of the fitted bands}

To understand the spin texture of the 2DEG, we have extended the TB calculation to compute the spin textures for all band pairs at different iso-energy lines. The spin textures for all the band diagrams shown in Fig. \ref{FS} are depicted in Fig. \ref{spin}. The size and direction of the arrows correspond to the magnitude and in-plane orientations of the spins. The color of the arrows indicates the out-of-plane component with red along \hkl[111], and blue along \hkl[-1-1-1]). In our previous study of KTO~(001) 2DEGs, we showed that near the band edge, each band pair has a k-linear Rashba-like splitting with almost circular iso-energy lines and perpendicular spin-momentum locking \cite{varotto_direct_2022}. On the contrary, for KTO~(111) 2DEGs, the spin textures exhibit perpendicular spin-momentum locking only at the pink and orange band pairs along the high symmetry directions, i.e. along $\Gamma$-K and $\Gamma$-M. The surface of KTO (111) has $C_{3v}$ symmetry with the mirror planes perpendicular to \hkl(-110), which is reflected in the iso-energy lines as well as the spin textures shown in Fig. \ref{spin}. The $C_{3v}$ symmetry forces the states at each mirror plane (states along $\Gamma$-M and equivalent directions) to have zero spin component parallel to the mirror plane. Hence, the spin texture is in-plane at the tips of the star-shaped pink band pair, but exhibits sizeable out-of-plane components in between. Further, the momentum splitting for each pair of sub-bands is not the same at every $k$ point. As previously determined in Ref. \cite{bruno_band_2019}, the splitting for the pink and green band pairs is larger along the $\Gamma - \text{K}$ direction than along the $\Gamma - \text{M}$ direction. From the ARPES data, we observe that the bands of neighboring band pairs are closer along the $\Gamma - \text{K}$ direction which gives rise to a situation of avoided crossing similar to that found in the STO~(001) band structure\cite{vaz_mapping_2019}. Therefore, the momentum splitting as well as the Rashba coupling constant is greater close to these points. This should motivate experiments to study the Edelstein effect in such systems. The $C_{3v}$ symmetry of the system allows an in-plane spin polarization perpendicular to an in-plane applied charge current. However, the magnitude of the spin polarization as well as its orientation with respect to the charge current does not depend on the direction of the applied charge current. Despite the nonzero out-of-plane components of the spin texture, the net current-induced spin polarization is expected to be in-plane, enforced by the $C_{3v}$ symmetry of the system.

\section{Conclusion}

In summary, we have visualized the electronic band structure of superconducting KTO~(111) 2DEGs using ARPES. The results show similarities with those reported on KTO~(111) surfaces with the following main features: (i) the Fermi surfaces of the bands from the outer pair have a pronounced anisotropic, star-like shape while the other bands have more isotropic Fermi surfaces (inside the measurement window); (ii) the spin textures strongly deviate from a standard Rashba model, showing considerable warping for the outer band pair and radial rather than orthoradial spins for some of the inner bands; (iii) the band splitting is moderate at the band bottoms but larger at the end of and in between the branches of the star. The anisotropy of the Fermi surfaces and spin textures echo that of the superconducting critical temperature and mobility and our results thus constrain the applicability of the models proposed to explain superconductivity in KTO 2DEGs \cite{liu_tunable_2023}. They also suggest that spin-charge conversion effects from the Rashba-Edelstein effect should be highly anisotropic and deviate from predictions using standard models \cite{vaz_determining_2020}, which could bring novel functionalities for spin-orbitronics.

\section*{Supplementary Material}

Supplementary material contains information regarding the transport measurement, x-ray photoelectric spectra (XPS) and low energy electron diffraction (LEED) for the substrate as well as the sample, the analysis related to energy distribution curves (EDC) and momentum distribution curves (MDC) and additional ARPES data measured using linear vertical (LV) polarization.

\begin{acknowledgments}
This work was supported by the ANR QUANTOP Project (ANR-19-CE470006) and the ERC Advanced grant “FRESCO” (No 833973). I.M. acknowledges support from the DFG under SFB TRR 227.

\end{acknowledgments}

\section*{Author Declarations}
\subsection*{Conflict of Interest}
The authors have no conflicts of interest to disclose.
\subsection*{Author Contributions}
\textbf{S. Mallik}: Investigation (lead); Writing - original draft (lead); Conceptualization (equal). \textbf{B. Göbel}: Theory model (lead); Writing - original draft (equal). \textbf{H. Witt}: Investigation (supporting); Writing - editing (equal). \textbf{L. M. Vicente-Arche}: Investigation (equal); Writing - editing (supporting). \textbf{S. Varotto}: Investigation (equal). \textbf{J. Bréhin}: Investigation (supporting). \textbf{G. Ménard}: Investigation (supporting). \textbf{G. Saïz}: Investigation (supporting). \textbf{D. Tamsaout}: Investigation (supporting). \textbf{A. F. Santander-Syro}: Investigation (supporting). \textbf{F. Fortuna}: Investigation (supporting). \textbf{F. Bertran}: Investigation (supporting). \textbf{P. Le Fèvre}: Investigation (supporting). \textbf{J. Rault}: Investigation (supporting). \textbf{I. Boventer}: Writing - review and editing (supporting). \textbf{I. Mertig}: Theory model (equal); Writing - review and editing (supporting). \textbf{A. Barthélémy}: Investigation (supporting). \textbf{N. Bergeal}: Investigation (supporting); Writing - review and editing (supporting). \textbf{A. Johansson}: Conceptualization (equal); Theory model (equal); Writing - review and editing (equal). \textbf{M. Bibes}: Conceptualization (lead); Project administration (lead); Writing - review and editing (equal).

\section*{Data Availability}
The data that support the findings of this study are available from the corresponding author upon reasonable request.

\section*{References}
\bibliography{ARPES-KTO111}

\end{document}